\documentclass[sigconf]{acmart}
\AtBeginDocument{%
  }

\copyrightyear{2026}
\acmYear{2026}
\setcopyright{none}
\setcctype{by}
\acmConference[FSE'26 IVR]{Montreal, QC, Canada}
\acmDOI{10.1145/3803437.3805589}

\usepackage{balance}

\begin{document}

\title{Treating Run-time Execution History as a First-Class Citizen: Co-Versioning Run-time Behavior alongside Code}

\author{Marcus Kessel}
\email{marcus.kessel@uni-mannheim.de}
\orcid{0000-0003-3088-2166}
\affiliation{%
  \institution{University of Mannheim}
  \city{Mannheim}
  \state{Baden-Württemberg}
  \country{Germany}
}


\begin{abstract}
  Behavioral Co-Versioning remains absent from mainstream practice: while developers routinely version source code with Git, they rarely persist and query how run-time behavior evolves across revisions.
This paper argues that this mismatch contributes to a blind spot in software evolution analysis and CI, where rich execution information is discarded and typically reduced to pass/fail outcomes --- despite partial test oracles, flakiness, and silent output or performance drift.
We propose \textit{Behavioral Co-Versioning}, a paradigm that couples the Git history with a \textit{Behavioral Archive}: an append-only, queryable store of selected run-time observations (e.g., method I/O and performance signals) collected during test runs and keyed by commit and test context.
This enables semantic diffing, behavior-aware regression localization, and retrospective auditing by querying historical executions, complementing proactive, signal-specific monitoring tools. We first outline a minimal data model and change diagnostics based on code/test/behavior fingerprints, and then demonstrate feasibility with a laptop-scale prototype that replays historical commits of a Python project, archives run-time observations in a local Parquet-backed store, and detects behavioral changes not apparent from textual diffs.
\end{abstract}




\maketitle

\section{Introduction}
\label{sec:introduction}

For years, approaches for Mining Software Repositories (MSR) have analyzed software evolution primarily through \emph{static} artifacts. Mining ASTs, code diffs, and commit messages yields deep insights into developer intent and structural change, but it is inherently limited in capturing \emph{dynamic semantics} (i.e., actual run-time behavior) due to Rice's Theorem \cite{riceTheorem,ernst2003static}. As a result, much of MSR (and, increasingly, Generative AI for SE trained on static corpora) models \emph{what code looks like} rather than \emph{what it does} at execution time \cite{NEURIPS2024_6efcc7fd,kessel2025}.

At the same time, modern Continuous Integration (CI) pipelines \cite{fowler_ci} spend substantial computing power executing test suites to generate rich run-time signals (e.g., outputs, traces, coverage, and timing). Yet standard practice discards most of these observations and reduces each run to a binary outcome: \textit{Pass} or \textit{Fail}. This reduction has well-known pitfalls: a passing suite can hide subtle output shifts \cite{5770598}, increasing flakiness \cite{10.1145/2635868.2635920}, or silent performance regressions \cite{5562942}. These issues are amplified by the oracle problem: many exercised behaviors remain unchecked due to weak or partial assertions \cite{barrOracle14,danglot2019automatic,taromirad2025assertions}. Consequently, CI often validates only what developers anticipated to assert, leaving other behavioral drift (i.e., actual run-time behavior changes) invisible.

This creates a fundamental blind spot: while code versioning (e.g., Git \cite{git-doc}) is mature, there is no widely-adopted analogue for persisting and querying behavior over time at the granularity of code units. This is striking because the benefits of code versioning are precisely the capabilities that engineers need for behavior: stable identifiers, reviewable change artifacts, reproducibility, bisection, and rollback. However, a commit hash is an imperfect proxy for behavior: run-time outcomes can change without source edits \cite{ammann2017introduction} (e.g., configuration/feature flags, dependency upgrades, nondeterminism, workload drift), and even source changes can yield behavioral consequences that a partial test oracle fails to detect.

We propose \textit{Behavioral Co-Versioning} (BeCoV): a paradigm that complements versioning code (as text) with versioning observed run-time behavior. The core idea is to couple the repository graph (e.g., Git commits) with a \textit{Behavioral Archive} that stores behavior snapshots produced during test executions (e.g., method I/O, selected state summaries, and performance signals) keyed by revision and test context. Whereas Git tracks changes in the definition of a code unit via textual diffs, BeCoV tracks changes in its manifestation via observational data. Aligning these histories --- linking code revisions to behavioral fingerprints --- enables evolution analysis that is sensitive to semantics, including detecting and characterizing behavioral discrepancies even when tests pass.

Concretely, BeCoV can be understood as a \emph{data differencing} problem --- analogous to how git diff compares two versions of code: given two structured datasets of behavioral observations---one per software version---compute a meaningful, structured behavioral diff that developers can act upon.
A BeCoV pipeline would
(i)~capture method-level inputs, outputs, and side effects during test execution;
(ii)~structure them at both method invocation and test granularity; and
(iii)~compute a behavioral diff identifying what changed, was added, or removed across revisions.
This dual-level design lets developers drill down: a coarse summary reveals \emph{where} behavior shifted, while fine-grained records reveal \emph{how}.

Existing tools in CI pipelines (e.g., unit testing, performance measurement, quality gates) are typically proactive and specific: developers must decide in advance what signals to collect (e.g., latency percentiles) and what properties to check. BeCoV is instead retrospective and generic: it preserves a reusable record of run-time observations so that new questions and new oracles can be evaluated later by querying historical runs. 

We argue that BeCoV becomes feasible due to modern storage and analytics techniques (e.g., columnar formats, compression, encodings) \cite{armbrust2021lakehouse,kessel2026observationlakehouseslivinginteractive,kesselOS2023}, which suggest that structured run-time observations can be stored efficiently as a repository-integrated, queryable history rather than as semi- or unstructured ephemeral logs.

Treating CI executions as historical data enables longitudinal queries that are currently impractical, such as: \textit{``How did the distribution of return values (or latency) of \texttt{calculate\_discount()} change over the last 50 commits?''}. This supports semantic diffing for review/refactoring, behavioral regression localization, and forensic auditing (re-checking historical executions against newly discovered constraints without re-running old revisions).

Realizing this vision raises research challenges around (1) \textit{volume} (capture/storage cost), (2) \textit{identity preservation} (linking co-evolving tests to exercised code units), and (3) \textit{representation/observability} (what to record and how to serialize it with acceptable overhead and then compare it). To ground feasibility, in this paper we present a minimal proof-of-concept on the \texttt{dateutil} Python library: we instrument a standard \texttt{pytest} suite and re-execute it across historical commits to populate a prototype behavioral archive linked to Git history, surfacing behavioral changes not apparent from static diffs alone. The remainder of this paper is structured as follows. Section~\ref{sec:core_problems} details the utility of BeCoV. Section~\ref{sec:co_versioning} outlines a preliminary model and a prototype demonstration. We conclude with reflections and a call for the community to treat run-time behavior as a first-class, versioned artifact of software evolution.
\section{The Case for Behavioral Co-Versioning}
\label{sec:core_problems}

Existing run-time verification tools (e.g., unit testing) are valuable, but proactive and specific: developers must decide in advance which signals to instrument and which properties to assert, validating only what was anticipated at development time.
BeCoV pursues a complementary goal---transforming CI executions from ephemeral checks into a queryable behavioral history---enabling retrospective analyses that are difficult to obtain from traditional CI artifacts.

We refer to the observable signals archived per code unit collectively as its \emph{behavioral fingerprint}: input/output values, internal call sequences, side-effect summaries, and performance characteristics recorded under the test suite.

\paragraph{Semantic Diffing for Review and Refactoring.}

Textual diffs are often a poor proxy for behavioral impact, especially for refactorings that restructure code without intending to change semantics~\cite{ALOMAR2021106675}.
By comparing behavioral fingerprints before and after a change, BeCoV can highlight which code units exhibit observable drift and filter syntactic noise during review.
Because portions of an execution trace are associated with multiple involved units, the archive also reveals cross-unit ripple effects, supporting queries such as: \textit{``Which downstream units exhibited behavioral drift after changes to \texttt{HelperUtil}?''}---a behavior-centric notion of impact largely invisible to assertion-local unit testing.

\paragraph{Behavior-Aware Regression Localization.}

For two consecutive green builds, methods may return different values for the same inputs, call sequences may be reordered, and new dependencies may be introduced --- all without any test failing.
Such silent behavioral changes arise whenever assertions cover only a subset of observable behavior (i.e., partial oracles \cite{barrOracle14}). For example, when a developer asserts only on the final return value of a method, or on a subset of an object's state, any change to intermediate computations, internal call sequences, or side effects may remain invisible to the test verdict.
A behavioral archive surfaces these shifts as fingerprint diffs even when CI remains green, and enables localization by comparing fingerprint distributions across revisions.
More broadly, project-level histories can yield implicit behavioral baselines (e.g., stable output schemata or performance envelopes) that complement explicit assertions and flag anomalous drift automatically.

\paragraph{Retrospective Auditing and Forensics.}

A behavioral archive enables post-hoc evaluation of properties that were not encoded as assertions at development time.
After a vulnerability report or a newly introduced compliance constraint, a team can query archived observations to assess when a problematic behavior first appeared and how broadly it manifested---without rebuilding and rerunning historical revisions, which is often impeded by dependency rot.

\paragraph{Downstream Opportunities.}

A standardized behavioral archive that continually grows (as proposed in \cite{kessel2025} for enabling Morescient GAI) also provides potential training and evaluation data for execution-aware developer tools (e.g., AI agents suggesting missing assertions from observed invariants, or synthesizing regression tests from historical input/output patterns and historically fragile boundary cases).
We view these as downstream beneficiaries of the archived information rather than its primary motivation.

\section{Model and Minimal Prototype}
\label{sec:co_versioning}

This section sketches a concrete realization of BeCoV and provides preliminary evidence via a minimal prototype study. The key idea is to couple the Git commit history with a behavioral archive (inspired by \cite{kessel2026observationlakehouseslivinginteractive}): an append-only collection of execution observations produced by CI/test runs and keyed by revision and test context. In contrast to CI artifacts that are typically ephemeral (pass/fail, raw logs), the archive treats executions as a persistent, structured dataset that supports longitudinal, behavior-aware queries. The prototype emphasizes \emph{end-to-end feasibility} (capture $\rightarrow$ store $\rightarrow$ query) and demonstrates behavior-aware change classification. Richer query models and validation of derived labels remain open for exploration.

\subsection{Conceptual Model}

BeCoV aligns two complementary histories:
(i) the \textit{code history} (the Git DAG), and
(ii) the \textit{behavior history} (execution records indexed by commit, test, and exercised code units).
A behavior record captures a selected set of observations from a test execution under a given revision (e.g., inputs/outputs at call boundaries, exceptions, and performance signals such as latency). Queries over these records enable behavior-centric views of evolution (e.g., drift, instability) that are not visible from textual diffs alone.

We model the archive as a table of records of the form --

\vspace{-0.2cm}
\[
\begin{aligned}
\langle{}& Commit\_ID,\; Test\_ID,\; Unit\_ID,\\
        & Test\_Hash,\; Unit\_Hash,\; Obs,\; Obs\_Hash,\;Context \rangle
\end{aligned}
\]

where \texttt{Obs} is a (potentially partial) serialized observation payload (e.g., method I/O and latency), and \texttt{Obs\_Hash} is a normalized fingerprint used for efficient comparison. \texttt{Test\_ID,Test\_Hash}, \texttt{Unit\_ID,Unit\_Hash} are the identified (test) code units and their hashes, and \texttt{Context} is the test context (e.g., environment). This design makes two assumptions explicit: (1) the archive captures observations under the test suite (per test procedure) in a specific context, not universal program semantics; and (2) determinism is not guaranteed \cite{4815280} --- hence both payloads and fingerprints may exhibit drift due to nondeterminism, environmental variation, or representation.

In the prototype, records are inspired based on the technical realization of the stimulus-response matrix (SRM) data structure proposed in \cite{kesselOS2023}. For each captured invocation, we serialize (i) the stimulus (inputs) and (ii) the response (return value/exception and timing) into JSON --- allowing for analytical queries over classic tabular representations. To enable longitudinal comparisons, we apply lightweight normalization (e.g., replacing execution-specific identifiers such as object instance IDs with stable placeholders like documented in \cite{softwareobservatorium_ssn_2025}). The observation fingerprint \texttt{Obs\_Hash} is then computed from this normalized representation; comparisons are performed using string equality.

\subsection{Ingestion: Capturing Observations}

A practical instantiation of BeCoV requires collecting observations with low friction and acceptable tracing overhead. In the prototype, we implement ingestion as a lightweight extension to \texttt{pytest} \cite{pytest_docs} (a popular unit testing framework for Python) that hooks into the test lifecycle and records a minimal observation schema: (i) call-boundary inputs/outputs for selected focal units, (ii) exceptions, and (iii) coarse-grained timing (latency) per invocation. The ingestion pipeline streams these observations into the behavioral archive together with the relevant code/test hashes.

\paragraph{Observability boundary}

The prototype intentionally adopts a \emph{minimal} observation tracing schema to reduce run-time overhead and data volume. Capturing deeper internal state (e.g., heap graphs) is possible in principle, but raises substantial representation and performance challenges. We treat the granularity of observation (and its efficient normalization) as a first-class research question rather than fixing it a priori. These open questions are addressed in our research roadmap in Section~\ref{sec:reflections}.

\subsection{Storage and Querying Feasibility}

Storing observations for every execution can be costly, but recent data-management techniques make persistent archival increasingly plausible. We adopt a data lakehouse layer \cite{armbrust2021lakehouse} for the behavioral archive, inspired by the ``observation lakehouse'' style persistence layer proposed in \cite{kessel2026observationlakehouseslivinginteractive,observationlakehouse2025}: observations are serialized into tables in terms of columnar files (using the Parquet format \cite{apache_parquet_format}), and partitioned \cite{duckdb_hive_partitioning} to support selective access to code units and their run-time behavior (avoiding full table scans). Columnar compression and encoding offered by columnar storage can reduce storage overhead when tests repeatedly yield identical or highly similar observations. Queries are executed directly over Parquet using an embedded analytical engine (DuckDB \cite{duckdb}), avoiding a dedicated server and enabling interactive analysis in developer-local settings.

\paragraph{Identity Preservation.}

Developers typically reason about tests, but BeCoV must attribute observations to the \emph{functional abstractions} those tests exercise (i.e., the code units that actually deliver the behavior under scrutiny). This attribution problem has two facets.

First, given a test execution, which code units constitute the \emph{focal units} of interest versus incidental infrastructure (logging, serialization, framework glue) (cf. \cite{7335402,9978960})? Heuristics such as package boundaries, or naming conventions offer starting points, but no single strategy is universally reliable.

Second, once focal units are identified, their identity
must be preserved across revisions (as in code versioning \cite{10.1145/3611643.3616312,10.1145/3236024.3264598}): methods are renamed, classes are split, and tests themselves co-evolve.
Without robust lineage tracking, behavioral diffs risk
comparing non-corresponding units and producing
misleading change reports.

\subsection{Behavior-aware Change Classification}
\label{sec:drift_categories}

Given successive revisions, we can compare code and observation fingerprints to obtain a lightweight diagnostic of how a unit and its tests co-evolve. For a fixed (\texttt{Test\_ID}, \texttt{Unit\_ID}), comparing revision $t$ to $t{-}1$ yields the following fundamental set of behavior-aware change categories:

\begin{itemize}
    \item \textbf{Observed behavior preserved} ($Test{=},\; Code\Delta,\; Obs{=}$): code changed but archived observations did not change under the chosen observation schema (candidate refactoring/optimization, or behavior not captured by the schema).
    \item \textbf{Observed behavioral drift} ($Test{=},\; Code\Delta,\; Obs\Delta$): code changed and observations changed (candidate regression or intended behavior change).
    \item \textbf{Instability / nondeterminism} ($Test{=},\; Code{=},\; Obs\Delta$): code did not change but observations changed (candidate flakiness, nondeterminism, environment drift, or representation noise).
    \item \textbf{Co-evolution} ($Test\Delta,\; Code\Delta$): both test and code changed, complicating direct drift attribution.
\end{itemize}

We emphasize that these categories are diagnostic heuristics over observed executions, not ground-truth labels or proofs of semantic equivalence --- hence they may serve as additional developer feedback sent to developers as part of CI. A key research direction is to develop robust query patterns and validation methodologies that distinguish true behavioral change from observational artifacts.

\paragraph{Minimal Feasibility Study.}

To ground feasibility, we implemented a minimal pipeline for the \texttt{dateutil} Python library \cite{dateutil2024}. A lightweight \texttt{pytest} extension captures method-level inputs, outputs, and latency for heuristically identified focal units, writing observation records to a local Parquet-backed columnar store with DuckDB. A SQL-based diff engine then compares behavioral snapshots across consecutive commits. The prototype is intentionally minimal: its purpose is to demonstrate that behavioral observations \emph{can} be captured and diffed within an existing test workflow, not to evaluate effectiveness at scale. A thorough empirical evaluation---including quantitative characterization of detected changes, storage overhead, and developer utility---is the subject of ongoing work. For this proof-of-concept study we partition by repository and fully qualified names of focal units (e.g., \texttt{dateutil.parser.parse}) and use heuristic attribution of tests to focal units. This design prioritizes simplicity and queryability; it does not implement robust lineage tracking across complex refactorings.

\paragraph{Preliminary observations.}

Across \texttt{dateutil}'s git history (past 100 commits successfully replayed; $28.136$ unique test units; $935$ focal code units), we populated a behavioral archive of size $\approx$131MiB (the focal unit \texttt{dateutil.parser.\_parser.parse} accounted for $\approx$104MiB), and executed the behavior-aware change classification (SQL query) in $434$ms on a commodity laptop. The query produced instances of the change categories. In this vision paper, we do not claim these instances are corresponding to ground-truth; they may also reflect environmental differences during replay or limitations of the observation representation. Further investigation with more controlled experimental conditions would be needed to determine whether the classifications are valid. Nevertheless, the result demonstrates the central premise of BeCoV: once execution observations are archived and keyed to commits, such hypotheses become \emph{queryable} and can be investigated systematically rather than being lost after CI completes.
For space reasons, we omit implementation details and additional query examples. The prototype, scripts, and datasets are available for inspection in the accompanying artifact \cite{kesselReplication2026}.
\section{Discussion, Related Work, and Conclusion}
\label{sec:reflections}

BeCoV sits between MSR (mining versioned static artifacts), dynamic analysis \cite{ball99,4815280}/regression/differential testing (collecting traces for immediate V\&V, including fault localization \cite{7390282}) \cite{mckeeman1998differential,ammann2017introduction}, and behavioral data infrastructure (e.g., lakehouse-style observation storage). Prior work provides datasets and monitoring techniques, but largely lacks a commit-aligned, queryable behavioral history for longitudinal tasks such as semantic diffing, behavior-aware regression localization, and retrospective auditing. BeCoV reframes CI executions as a persistent data asset rather than an ephemeral (quality) gate or guardrail. This shift comes with limitations and trade-offs that delimit the vision and suggest a research agenda.

\paragraph{Limitations and Trade-offs}

\textbf{Coverage.} BeCoV is inherently based on execution: code that is not exercised (by tests or probes) has no behavioral history. In practice, many projects maintain substantial automated test suites, yet test coverage is often incomplete and unevenly distributed across code units \cite{10.1145/2568225.2568271}. We therefore view BeCoV as complementary to static techniques, and as increasingly viable as automated test generation and probing reduce uncovered code units. \textbf{Economics.} At first glance, BeCoV appears to conflict with an industry trend toward \emph{test reduction} (selection/prioritization/minimization \cite{10.1145/2635868.2635910}) to save CI time and compute. We argue that BeCoV is orthogonal: even when the same reduced set of tests is executed, retaining selected execution observations can amortize its cost by enabling post-hoc queries (e.g., auditing, regression localization) without repeated re-execution and manual reproduction. In this sense, BeCoV shifts effort from repeated \emph{compute-time} and \emph{developer-time} expenditures toward a controlled, explicit \emph{data retention} cost. \textbf{Observability.} Capturing (tracing) ``more behavior'' increases overhead. A practical BeCoV system must support configurable observation schemata (e.g., I/O summaries, latency, exceptions) and normalization mechanisms, while acknowledging that archived observations represent behavior \emph{under recorded test contexts}, not universal program semantics. \textbf{Nondeterminism.} Behavioral drift may reflect nondeterminism (at value and sequence level), dependency changes, or platform variation rather than source edits. BeCoV does not eliminate these factors; instead, it makes them measurable and queryable, enabling explicit analysis of instability.

\paragraph{Scalability Challenges.}

In large-scale systems with large test sets and frequent daily commits, naive capture-everything strategies are likely infeasible. Three design dimensions shape the storage--precision trade-off:
\emph{observation depth} (how deep into the call chain to trace---shallow observation reduces volume, but may miss transitive changes---finding a balance between tracing too much vs. too little);
\emph{change-aware indexing} (e.g., structural fingerprinting to avoid revisiting unchanged portions before diffing); and
\emph{serialization profiles} (normalizing non-deterministic input/output values such as timestamps or memory addresses that otherwise introduce diff noise). On the infrastructure side, the lakehouse-style architecture described in Section \ref{sec:co_versioning} naturally maps onto cloud object stores (e.g., Amazon S3 storage \cite{armbrust2021lakehouse,aws_s3_lakehouse_2025}); columnar Parquet files can be directly read from and written to such stores at negligible marginal cost, making long-term retention of behavioral archives economically viable even for large mono-repositories.
Additionally, practical deployment requires lifecycle policies for the behavioral archive---e.g., retention windows, incremental snapshot updates, and integration with existing CI storage budgets---that remain open engineering challenges.
Each dimension presents an open trade-off between precision, cost, and generality that future work must investigate.

\paragraph{Conclusion}

With Behavioral Co-Versioning (BeCoV) and a Behavioral Archive, we have argued that run-time behavior deserves the same versioning discipline that source code receives today.
Our minimal prototype suggests basic feasibility with off-the-shelf instrumentation and lakehouse-style storage, but making BeCoV as routine as source code versioning requires progress on several fronts:
(1)~\emph{Identity preservation}---attributing observations to functional abstractions and tracking lineage across refactorings and test co-evolution;
(2)~\emph{Representation}---designing standardized behavioral snapshot schemata, robust fingerprinting under nondeterminism (e.g., via serialization profiles), and choosing appropriate observation tracing depth;
(3)~\emph{Scalable indexing}---leveraging change-aware fingerprinting structures to further improve storage and comparison;
(4)~\emph{Workflow integration}---low-friction capture and querying in CI/IDE settings, with clear cost controls; and
(5)~\emph{Branching and merging}---behavioral diffing techniques to identify branch-specific behavioral changes and potential conflicts at merge time, analogous to three-way textual merge in Git (i.e., comparing each branch's files with the ancestor to detect conflicting edits).

By surfacing behavioral changes that current CI pipelines miss and by opening avenues such as behavioral diffing for software evolution, BeCoV offers a complementary lens to the purely textual code view of software history that dominates current practice.

\balance
\bibliographystyle{ACM-Reference-Format}
\bibliography{literature}

\end{document}